\documentclass[fdp,10pt]{w-art}
\usepackage{amssymb}
\usepackage{graphicx}
\usepackage{dcolumn}
\usepackage{bm}

\newcommand\MR{\mathbb{R}}
\newcommand\bfp{\mathbf{p}}
\newcommand\bfpp{\mathbf{p}'}
\newcommand\bfx{\mathbf{x}}
\newcommand\bfu{\mathbf{u}}
\newcommand\bfv{\mathbf{v}}
\newcommand\bft{\mathbf{t}}
\newcommand\bfta{\mathbf{t}_a}
\newcommand\bfk{{\mathbf{k}_a}}
\newcommand\bfn[1]{\mathbf{n}_#1}
\newcommand{\sgn}{\operatorname{{\mathrm sgn}}} 
\newcommand{\tr}{\operatorname{{\mathrm tr}}}

\begin{document}
\def\MR{{\mathbb{R}}}


\title[Fermions Violate Area Law]{Free Fermions Violate the Area Law For 
Entanglement Entropy}

\keywords{Entanglement entropy, Area Law, LMU-ASC-08-10}

\author[RC Helling]{Robert
  C. Helling\inst{1,}
}
\address[\inst{1}]{Arnold Sommerfeld Center,
  Ludwig-Maximilians-Universit\"at M\"unchen\\
} 
\author[W Spitzer]{Wolfgang
  Spitzer\inst{2,}
}
\address[\inst{2}]{
  Institut f\"ur Theoretische Physik,
  Universit\"at Erlangen-N\"urnberg\\
}
\date{\today}

\begin{abstract}
We show that the entanglement entropy associated to a region grows
faster than the area of its boundary surface. This is done by
proving a special case of a conjecture due to Widom that yields a
surprisingly simple expression for the leading behaviour of the
entanglement entropy. 
\end{abstract}

\maketitle

\section{Background}

An interesting model for the area law of Bekenstein's black-hole
entropy is a local quantum field theory for which one restricts all
observations to the complement of a compact spatial region $\Omega$. One 
should think of $\Omega$ as the interior of the horizon although one does 
not require $\Omega$ to have any gravitational relevance. In fact, we will 
restrict our attention to a region in the $n$-dimensional flat Euclidean space
$\MR^n$.  Even if the quantum field is globally in a pure state
(e.g., the vacuum), the state restricted to the complement of $\Omega$ will 
in general be mixed with finite von Neumann entropy, called the 
{\em entanglement entropy}, $S(\Omega)$~\cite{Sred,BKLS,eisert}. 
For many types of quantum field theories, it was observed 
that in the semi-classical limit of large $\Omega$, the entanglement entropy
$S(\Omega)$ scales like the area of $\partial\Omega$. In other words, if we
rescale $\Omega\subset\MR^n$ by a factor $R$, the entanglement entropy, 
$S(R\Omega)$, should asymptotically scale like $R^{n-1}$ for large $R$.

Recently, it was observed by Gioev and Klich~\cite{Gioev,GK}, that for free
fermions at zero temperature, however, the entanglement entropy of the
ground state scales like $R^{n-1}\log(R)$ 
for $R\to\infty$ based on a conjecture by Widom~\cite{Widom,Widom2}. They
provide a lower bound on the entanglement entropy in terms of the trace of 
a quadratic function of the restricted state. For the latter, we have proved 
in \cite{HLS} the leading asymptotic behaviour. This establishes indeed that the
entanglement entropy scales at least like $R^{n-1}\log(R)$, which 
violates of the beforementioned area law scaling. 

For notational clarity we will often write equalities
that only hold asymptotically for large $R$, that is, we will drop
subleading terms that are not central to our argument and only mention
it in the text. 

General arguments imply that for a pure state, the entanglement entropy 
of this state with respect to $\Omega$ is the same as for the complement 
$\MR^n\setminus \Omega$. For simplicity, we will restrict the state
to a compact region $\Omega$.  The ground state, $\rho_\Gamma$, of a system 
of non-interacting fermions in $\MR^n$ is given in terms of the Fermi surface 
at Fermi energy $\epsilon_F$: All one-particle states with momentum
$\bfp\in\Gamma=\{\bfp\in\MR^n|E(\bfp)\le\epsilon_F\}$ and energy\footnotemark $E(\bfp)$ are 
occupied.
\footnotetext{We will not discuss here assumptions on the dispersion relation
$E(\bfp)$ but one may, of course, think of the example $E(\bfp)=\bfp^2$. We
require that $\Gamma$ and $\Omega$ are compact sets in $\MR^n$ with 
sufficiently smooth boundaries.} 
The ground state $\rho_\Gamma$ is then characterised by the one-particle Fermi
projector, $P_\Gamma$, defined in momentum space by the kernel 
$P_\Gamma(\bfp,\bfpp)=\chi_\Gamma(\bfp)\delta(\bfp-\bfpp)$. Here, and in the 
following, $\chi_A$ denotes the indicator function of a set $A$. In position
space, the kernel $P_\Gamma(\bfx,\bfx') = \widehat{\chi_\Gamma}
(\bfx-\bfx')$ with $\widehat{\chi_\Gamma}(\bfx)=(2\pi)^{-n}\int_\Gamma d\bfp \,
e^{i \bfx\cdot\bfp}$ being the inverse Fourier transform of $\chi_\Gamma$.

In order to restrict the state $\rho_\Gamma$ to the region $\Omega$ we project 
the Fermi projector $P_\Gamma$ onto $L^2(\Omega)$ with $Q_\Omega=\chi_\Omega$. 
This gives the reduced one-particle density matrix $\varrho_{\Omega,\Gamma}=
Q_\Omega P_\Gamma Q_\Omega$. The entanglement entropy, $S(\Omega,\Gamma)$, 
of the many particle system in the ground state $\rho_\Gamma$ restricted to 
the region $\Omega$ is then defined as the grand canonical entropy of 
$\varrho_{\Omega,\Gamma}$, that is, $S(\Omega,\Gamma) = \tr 
\eta(\varrho_{\Omega,\Gamma})$ with $\eta(t) = -t\log{(t)} -(1-t)\log{(1-t)}$
for $0<t<1$. For details see \cite[Section 4]{HLS}. We are interested here in 
the behaviour of this entropy for fixed $\Gamma$ but large $\Omega$. To 
this end, we also fix $\Omega$ and study the asymptotic behaviour of 
$S(R\Omega,\Gamma)$ as $R\to\infty$.

Our main result is the asymptotic computation of $\tr[\varrho_{R\Omega,\Gamma} 
(1 - \varrho_{R\Omega,\Gamma})]$ as $R\to\infty$:
 \begin{eqnarray}\label{quadwid}
    \label{eq:bound}
    &&\tr[\varrho_{R\Omega,\Gamma} (1 - \varrho_{R\Omega,\Gamma})]
    =\left(\frac{R}{2\pi}\right)^{n-1}\frac{\log(R)}{4\pi^2}
    \int_{\partial\Omega\times\partial\Gamma}
    dA(\bfx)dA(\bfp)\,|\bfn\bfx\cdot\bfn\bfp| ,
  \end{eqnarray}
up to terms that grow slower in $R$. Here, $\bfn\bfx$ denotes the
unit normal vector at $\bfx\in\partial\Omega$, $dA(\bfx)$ is the surface 
measure on $\partial\Omega$, and similarly for $\bfn\bfp$ and $dA(\bfp)$.

Since $\eta(t)\ge\log(2) 4 t (1-t)$ for $0<t<1$, we
obtain the asymptotic lower bound on the entanglement entropy, 
\begin{equation}
\label{S:lower bound}
S(R\Omega,\Gamma) \ge\frac{\log(2)}{\pi^2} \left(\frac{R}{2\pi}
\right)^{n-1}\log(R)\int_{\partial\Omega\times\partial\Gamma}
dA(\bfx)dA(\bfp)\,|\bfn\bfx\cdot\bfn\bfp|.
\end{equation} 
Gioev and Klich conjectured in \cite{GK} that the exact scaling of 
$S(R\Omega,\Gamma)$ is obtained if we replace the factor $\log{(2)}/\pi^2$ in 
\eqref{S:lower bound} by $\frac{1}{12}$. This remains an open problem.

\section{Computation of $\tr[\varrho_{R\Omega,\Gamma} (1 -
\varrho_{R\Omega,\Gamma})]$}

The trace of $\varrho_{R\Omega,\Gamma}$ is simply equal to 
$\left(\frac{R}{2\pi}\right)^n|\Omega||\Gamma|$, where $|\cdot|$ denotes the
$n$-dimensional Lebesgue volume. The trace of $\varrho_{R\Omega,\Gamma}^2$
equals
\begin{eqnarray}
  \label{eq:rhosquared}
  \tr(Q_{R\Omega}P_\Gamma Q_{R\Omega}P_\Gamma)\!\!\! 
&=&
\!\!\!\int_{R\Omega}\!\!\! d\bfx\int_{R\Omega}\!\!\! d\bfx'\,|\widehat{\chi_\Gamma}(\bfx-\bfx')|^2
=
\int_{R(\Omega-\Omega)}\!\!\! d\bfv \,|\widehat{\chi_\Gamma}(\bfv)|^2 
\,|R\Omega\cap (R\Omega-\bfv)|\,.
\end{eqnarray}
In the last line we have changed the variables $\bfx$ and $\bfx'$ to
$\bfu=\bfx$ and $\bfv=\bfx-\bfx'$. Then we expand the volume 
$|R\Omega\cap (R\Omega-\bfv)|$ to first order in $\bfv$ (cf. 
\cite[Theorem 2.1]{Roccaforte}),
\begin{equation}
  \label{eq:intersectvolume}
  |R\Omega\cap (R\Omega-\bfv)| = R^n|\Omega|+R^{n-1}\!\!\!
  \int_{\partial\Omega}\!\!\!dA(\bfx)\, \max(0,\bfv\cdot \bfn\bfx)+
  R^{n-2}O(|\bfv|^2). 
\end{equation}
Let us first look at the contribution of $R^n|\Omega|$ to the trace of 
$\varrho_{R\Omega,\Gamma}^2$. The function $\bfv\mapsto\widehat{\chi_\Gamma}(\bfv)$ decays 
like $|\bfv|^{-(n+1)/2}$ for large $|\bfv|$, see \eqref{eq:firststdesc}.
At the cost of an order $R^{n-1}$-term we may therefore extend the 
$\bfv$-integration to all of $\MR^n$. By the Plancherel formula this integral
gives $(2\pi)^{-n}|\Gamma|$ and cancels with $\tr\varrho_{R\Omega,\Gamma}$.
When integrated over $\bfv$, the remainder term $R^{n-2}O(|\bfv|^2)$ is also 
seen to yield a term of the order $R^{n-1}$ by using again the above
mentioned decay of $\widehat{\chi_\Gamma}$. 

Thus, our bound on the entanglement entropy will come from the second term 
in (\ref{eq:intersectvolume}). Here, we write $\max(0,\bfv\cdot \bfn\bfx)=
\chi_{[0,\infty)}(\bfv\cdot \bfn\bfx)\,\bfv\cdot \bfn\bfx$ and use the Gau\ss {}
 Theorem so that 
\begin{equation}\label{eq:pint}
  (2\pi)^n \,\bfv \,\widehat{\chi_\Gamma}(\bfv) =
  -i\int_{\partial\Gamma}dA(\bfp)\,\bfn\bfp \,e^{i\bfv\cdot\bfp}.
\end{equation}
So it remains to show that for $\bfp\in\partial\Gamma$,
\begin{equation}\label{22}
\Big|\int_{R(\Omega-\Omega)} d\bfv\, \chi_{[0,\infty)}(\bfv\cdot \bfn\bfx)
\,\widehat{\chi_\Gamma}(-\bfv) \,e^{i\bfv\cdot\bfp} + (2\pi i)^{-1} 
\sgn({\boldsymbol n}_{\bfx}\cdot {\boldsymbol n}_{\bfp})\log{(R)}\Big| 
= o(R)\,.
\end{equation}
Let us now consider the function $\bfv\mapsto\widehat{\chi_\Gamma}(-\bfv)$ 
in detail. We use the representation from \eqref{eq:pint}, that is,
$(2\pi)^{-n}\widehat{\chi_\Gamma}(-\bfv) = \frac{i \bfv}{|\bfv|^2}\cdot
\int_{\partial\Gamma}dA(\bfpp)\,\bfpp \,e^{-i\bfv\cdot\bfpp}$.
Then we introduce a coordinate system where $\bfv = (0,\ldots,0,v)$ and
where the boundary $\partial\Gamma$ is locally written as the graph of a
function $f\colon U\subset \MR^{n-1}\to \MR$, that is, $\bfpp=(\bft,f(\bft))$ 
and $dA(\bfpp) = \sqrt{1+|\nabla f|^2}d\bft$. The unit normal vector is
${\boldsymbol n}_{\boldsymbol p'} = \sgn(\bfv\cdot\bfpp) (-\nabla f,1)/
{\sqrt{1+|\nabla f|}}$.
Then, 
$$\int_{f(U)} dA(\bfpp)\, ({\boldsymbol n}_{\boldsymbol p'})_n\,
e^{-i\bfv\cdot\bfp}  = -\frac 1v\int_Ud\bft\,\sgn(f(\bft))\,e^{-iv f(\bft)}.
$$ 
In order to find the leading 
asymptotic behaviour of this $\bft$-integral for large $v$ we apply the method 
of stationary phase. Let $\bfk=\bfk(\bfv) = (\bfta(\bfv),f(\bfta(\bfv)))$ be 
the 
collection of all stationary points of such local functions $f$, that is, 
$\nabla f(\bfta)=0$; in other words, the points $\bfk\in\partial\Gamma$ are 
such that the unit normal vector $\boldsymbol n_{\bfk}$ at $\bfk$ is parallel 
to $\bfv$. Thus, 
\begin{eqnarray}
  \label{eq:firststdesc}
  \widehat{\chi_\Gamma}(-\bfv) 
  =-i (2\pi v)^{-(n+1)/2} \sum_{\bfk}
  \frac{\sgn(\bfv\cdot\bfk)}{\sqrt{|\det f_{ij}(\bft_a)}|}\, 
  e^{-i\bfv\cdot\bfk -i \frac\pi4 \sgn(f_{ij}(\bft_a))} + o(v^{-(n+1)/2}).
\end{eqnarray}
Here, $f_{ij}(\bft_a)$ denotes the Hessian of $f$ at $\bft_a$, 
$\sgn(f_{ij}(\bft_a))$ the signum of this Hessian. The determinant, 
$\det f_{ij}(\bft_a)$, equals the Gau\ss ian curvature of $\partial\Gamma$
at $\bfk$. 


Using \eqref{eq:firststdesc} we return to the oscillatory integral in 
\eqref{22}, and employ once more the method of stationary phase.  
The composite phase from (\ref{eq:pint}) and (\ref{eq:firststdesc}) is equal to 
$\bfv\cdot(\bfp-\bfk(\bfv))$. Next, we introduce generalised spherical 
coordinates for $\bfv$ as $\bfv=\rho(\bfu,h(\bfu))$, where the map $h\colon V
\subset\MR^{n-1}\to\MR$ locally parametrises the boundary $\partial(\Omega-
\Omega)\ni(\bfu,h(\bfu))$ and $\rho\in[0,R]$ is a radial coordinate. The
stationary point $\bfk$ is now a function of $\bfu$. We have the
freedom to assume that $\bfn\bfp=(0,\ldots,0,1)$ such that
$\bfk((0,h(0)))=\bfp$ for one index $a$ and that $h$ has its extremum at the
origin $\bfu=0$. Instead of integrating $\rho$ from 0 to $R$ we will actually
integrate $\rho$ only over $[1,R]$ thereby making an irrelevant error
independent of $R$.

To express $\bfk(\bfv)$ 
as a function of $\bfu$ we equate $\bfv/v=\boldsymbol n_{{\boldsymbol k}_a(\bfv)}$, 
that is,
\begin{equation}
  \label{eq:normals}
  \frac{(\bfu,h(\bfu))}{\sqrt{\bfu^2 + h(\bfu)^2}} =
  \sgn(\bfv\cdot\bfn\bfk) \frac{(-\nabla_\bft
    f(\bft),1)}{\sqrt{1+|\nabla_\bft f(\bft)|^2}}.
\end{equation}
Taking derivatives and evaluating at $\bfu=0$, we find $\frac{\partial t_j}{
\partial u_i}(0) = -f_{ij}^{-1}(\bfk(0))/h(0)$, where $\bfk(0)$ is short for 
$\bfk(0,h(0))$ and $f_{ij}^{-1}$ is the matrix inverse of the 
Hessian of $f$. With this, we can expand the phase of the remaining 
$\bfv$-integral to second order as 
\begin{equation}
  \label{eq:phaseuu}
  \bfv\cdot(\bfp-\bfk(\bfv)) = \rho h(0)(\bfp-\bfk(0))_n+ \rho
  \frac{f_{ij}^{-1}(\bfk(0))}{2h(0)}u_i u_j, 
\end{equation}
The volume element is given by $d\bfv = \rho^{n-1}h(\bfu)d\rho
d\bfu$. The stationary phase integral over the $n-1$ coordinates
$\bfu$ yields $(2h(0)/\rho)^{(n-1)/2} \sqrt{|\det f_{ij}^{-1}|}
\exp(i\rho h(0) (\bfp-\bfk(0))_n) + i \frac\pi4 \sgn(f_{ij}(\bft_a))$.
This, surprisingly, reduces the remaining $\rho$-integral to $\int_1^R d\rho\; 
e^{i\rho h(0) (\bfp-\bfk(0))_n}/\rho$. As noted above, for one index $a$, the 
exponent vanishes as $\bfp=\bfk(0)$ and the integral is the desired $\log(R)$. 
If, however, $\bfp\ne\bfk(0)$ the integral is bounded for large $R$ and does 
not contribute to leading order.

Collecting all terms we have thus proved the lower bound (\ref{eq:bound}) 
which grows faster than the area law scaling $R^{n-1}$
by a factor of $\log(R)$. Note that the stationary phase integrals
localises $\bfv$ and $\bfpp$ such that there are only contributions
from $\bfp$, $\bfn\bfv$, ${\boldsymbol n}_{\boldsymbol p'}$ all being parallel. 

\section{Discussion}
Above, we presented a lower bound for the entanglement entropy which violates 
an $R^{n-1}$ area law scaling. The expression (\ref{S:lower bound}) is 
in terms of integrals over the boundaries $\partial\Omega$ and $\partial\Gamma$ 
and is thus still a ``boundary effect''. 

We find it most curious that although we have used stationary phase methods, 
which in general depend on (the existence of) second order derivatives
at the stationary points, all these curvature terms involving $f_{ij}$ and
$h_{ij}$ eventually cancel out and the integrand in the bound does not
contain derivatives. However, one expects \cite{Gioev,fannes} that a fractal 
boundary $\partial \Omega$ of dimension $n-1+\alpha$ with $0<\alpha<1$ leads 
to a scaling of the entanglement entropy of at least of the order 
$R^{n-1+\alpha}$ and, presumably, without a $\log{(R)}$ correction. 

For $n=1$, if $\Omega$ and $\Gamma$ are a disjoint union of $k$ and $\ell$ compact intervals 
of finite length respectively, then our lower bound (\ref{S:lower bound})
for the entanglement entropy gives $\log(R) \log(2) 4k\ell/\pi^2$. In this
one-dimensional case, the precise scaling has been proved, namely, $\log(R) 
k\ell/3$. If we then consider the hypercubes, say $\Omega=\Gamma=[0,1]^n$, it is 
not difficult to derive the exact asymptotic scaling of $S(R\Omega,\Gamma)$ to 
be $(R/(2\pi))^{n-1} \log{(R)} n^2/3$, which is in agreement with the 
conjecture by Gioev and Klich mentioned at the end of Section 1. Our method of 
proof requires that the surfaces $\partial\Omega$ and $\partial\Gamma$ are
$C^3$. Hence, hypercubes are not included. For a $C^3$ surface, $\partial
\Gamma$, it was crucial that $\widehat{\chi_\Gamma}(\bfv)$ behaves like
$|\bfv|^{-(n+1)/2}$ for large $|\bfv|$. This is not the case for a non-smooth
surface such as the hypercube, where $\widehat{\chi_{[0,1]^n}}(\bfv) = 
\prod_{i=1}^n \sin(v_i)/(\pi v_i)$.

It should be noted as well that the discontinuity of $\chi_\Gamma$ is
crucial for the decay of its Fourier transform. For example, for the
equilibrium state at positive temperature, the entanglement entropy
scales like the volume $R^n$, see \cite{GK}. 
We did not have to introduce an ultraviolet regulator since
momentum integrations are limited to the compact region
$\Gamma$.



\section*{Acknowledgments}
We are grateful to Hajo Leschke with whom the work in \cite{HLS} was
performed on which is text is based.  Furthermore, RCH would like to
thank the Elitenetwork of Bavaria for financial support and Jacobs
University Bremen, where this work was started. We also thank Urs
Frauenfelder for discussions.

\bibliographystyle{fdp}
\bibliography{widom}

\end{document}